\documentclass[final,1p,times]{elsarticle} 
\usepackage{graphicx} 
\usepackage{amssymb} 
\usepackage{amsthm} 
\usepackage{lineno} 

\journal{Nuclear Physics A} 
\begin{document} 

\providecommand{\tabularnewline}{\\}

\newcommand{\pt}{ p_{\rm T}}
\newcommand{\ie}{{\it i.e.}}

\begin{frontmatter} 


\title{Elliptic flow of thermal photons at midrapidity in Au+Au collisions
at $\sqrt{s_{NN}}=200$ GeV}

\author{Fu-Ming Liu$^{a}$, Tetsufumi Hirano$^{b}$, Klaus Werner$^{c}$, Yan Zhu$^{a}$ }

\address[a]{Institute of Particle Physics, Central China Normal University, 430079, Wuhan, China}
\address[b]{Department of Physics, The University of Tokyo, 113-0033, Japan}
\address[c]{Laboratoire SUBATECH, University of Nantes - IN2P3/CNRS - Ecole desMines,
Nantes, France}

\begin{abstract} 
The elliptic flow $v_{2}$ of thermal photons at midrapidity 
in Au+Au collisions at $\sqrt{s_{NN}}=200$ GeV is predicted, 
based on three-dimensional ideal hydrodynamics.
Because of the interplay between the asymmetry and the strength 
of the transverse flow, the thermal photon $v_{2}$ reaches a maximum at  
$\pt \sim $  2GeV/$c$ and the  $\pt$-integrated $v_{2}$ reaches a
 maximum at about 50\% centrality. 
The $\pt$-integrated $v_{2}$ is very sensitive to the lower limit
 of the integral but not sensitive to the upper limit due to the
  rapid decrease in the spectrum of the transverse momentum.  
\end{abstract} 

\end{frontmatter} 



\section{Introduction }

The deconfined and novel nuclear matter, the quark gluon plasma (QGP),
has been expected to appear in relativistic heavy ion collisions.
The observation of large elliptic flow of different hadronic species at the 
Relativistic Heavy Ion Collider (RHIC) at Brookhaven National Lab, New York, 
is of special importance to confirm the the formation of QGP. Unlike those bulk hadrons, photons are produced during the whole history of the evolution of the hot
and dense matter. Moreover, the mean free path of photons is much larger
than the transverse size of the bulk matter. So the produced photons  pass through the surrounding matter without any interaction. As a result, thermal photons provide undistorted information on flow asymmetries not only from its surface but also from the inner of the hot, dense matter.

Therefore the question such as what is the relation between the measurable elliptic flow $v_{2}$ of thermal photons and the evolution process of the expanding hot dense matter is of interest. This relation will serve as a direct bridge between the observables and the properties of the quark gluon plasma. 
A pioneering work has been done based on 2D ideal hydrodynamics, which
has shown the transverse momentum and centrality dependence of the
elliptic flow of thermal photons at midrapidity\cite{ther}. 
In this paper, the elliptic flow of thermal photons is calculated 
based on 3D ideal hydrodynamics. The paper is organized as following.
In Sec.~2 we will briefly review the space-time evolution of the
hot and dense matter using 3D ideal hydrodynamics and the basic
formula for the production of thermal photons. In Sec.~3, we will
show our results on the transverse momentum and centrality
dependences of thermal photons $v_{2}$ in Au+Au collisions
at $\sqrt{s_{NN}}=200$ GeV. Section 4 is devoted to discussion and summary
of our results. 

\section{Bulk evolution, thermal photons, and elliptic flow}

In our calculation, a full 3D ideal hydrodynamic calculation 
is employed to describe the space-time evolution of the hot and dense
matter created in Au+Au collisions at RHIC. 
The impact parameters corresponding to different centralities in Au+Au
collisions at RHIC, which are estimated with a Glauber model, are 3.2,
5.5, 7.2, 8.5, 9.7, 10.8, and 11.7 fm for 0-10\%, 10-20\%, $\cdots$\,
and 60-70\% centrality, respectively. The local thermal equilibrium
is assumed to be reached at the initial time $\tau_{0}=0.6$ fm/$c$.
The critical temperature of a first order phase transition between
the QGP phase and the hadron phase is fixed at $T_{c}=170$ MeV. 
The initial flow is taken to be Bjorken scaling solution. So the transverse
flow is vanishing at $\tau_{0}$ and dynamically generated by the
pressure gradient. More details can be found in \cite{hyd1,hyd2}. 

Transverse momentum spectra of thermal photons can be written as \begin{equation}
\frac{dN}{dy\, d^{2}p_{t}}=\int d^{4}x\,\Gamma(E^{*},T)\label{eq1}\end{equation}
 with $\Gamma(E^{*},T)$ being the Lorentz invariant thermal photons
emission rate, $d^{4}x=\tau\, d\tau\, dx\, dy\, d\eta$
being the volume-element, the integration is done from the initial
time $\tau_{0}$ to the freeze-out, and $E^{*}=p^{\mu}u_{\mu}$ the
photon energy in the local rest frame. More details can be found in
\cite{Liu:2008eh}.

The elliptic flow is quantified by the second harmonic coefficient $v_{2}$ 
\begin{eqnarray}
v_{2}(\pt,y)=\frac{\int d\phi\cos(2\phi)d^{3}N/dy\, d^{2}p_{t}}{\int d\phi d^{3}N/dy\, d^{2}p_{t}},\end{eqnarray}
where $\phi$ is the azimuthal angle of photon's momentum with respect
to the reaction plane.  

The $p_{t}$ dependence of the triple differential spectra is strongly
affected by the flow $u$ through the argument $E^{*}=p^{\mu}u_{\mu}$
in the photon emission rate. The azimuthal asymmetry of the transverse
components of the flow obviously results in an anisotropic momentum
distribution, which gives a finite $v_{2}$.  Therefore both strength
and the anisotropy of transverse flow velocity are important to generate
the elliptic flow of thermal photons.
The two key features of QGP,
mean radial flow $\left\langle v_{r}\right\rangle $ 
and the mean anisotropy of flow $\left\langle v_{2}^{\mathrm{hydro}}\right\rangle $,
are defined as \begin{equation}
\left\langle v_{r}\right\rangle  = \left\langle \sqrt{v_{x}^{2}+v_{y}^{2}}\right\rangle,  
\left\langle v_{2}^{\mathrm{hydro}}\right\rangle  =\left\langle \frac{v_{x}^{2}-v_{y}^{2}}{v_{x}^{2}+v_{y}^{2}}\right\rangle ,\label{eq:aveflow2}\end{equation}
 where $\left\langle \cdots\right\rangle $ stands for energy-density-weighted
space-time average, $v_{x}$ and $v_{y}$ are the flow velocity components
along $x$-axis and $y$-axis, respectively. 
From Table~\ref{cent}, we see that the mean radial flow increases
with centrality from 0-10\% ($b=3.2$ fm) to 20-30\% ($b=7.2$ fm),
then decreases from 20-30\% ($b=7.2$ fm) to 60-70\% ($b=11.7$ fm).
On the other hand, the mean flow anisotropy increases with centrality
from 0-10\% to 60-70\% monotonically. The increase of the average
radial flow between 0-10\% and 20-30\% can be understood due to the
fact that the initial shape of matter becomes more asymmetric, leading
to flow asymmetry, as discussed earlier. The decrease beyond 20-30\%
is due to the decreasing life times, which prevents large flows from
developing.

\begin{table}[h]

\caption{\label{cent}The strength and the anisotropy of transverse flow at
each centrality.}  
 
 \begin{tabular}{cccccccc}
Centrality(\%) &
0-10 &
10-20 &
20-30 &
30-40 &
40-50 &
50-60 &
60-70\tabularnewline
$b$(fm) &
3.2 &
5.5 &
7.2 &
8.5 &
9.7 &
10.8 &
11.7\tabularnewline
$\left\langle v_{r}\right\rangle $ &
0.114 &
0.122 &
0.123 &
0.117 &
0.109 &
0.0959 &
0.0804\tabularnewline
$\left\langle v_{2}^{\mathrm{hydro}}\right\rangle $ &
0.0417 &
0.103 &
0.154 &
0.188 &
0.212 &
0.222 &
0.240 \tabularnewline
\end{tabular}
\end{table}

\section{Results}
\begin{figure}
\includegraphics[scale=0.8]{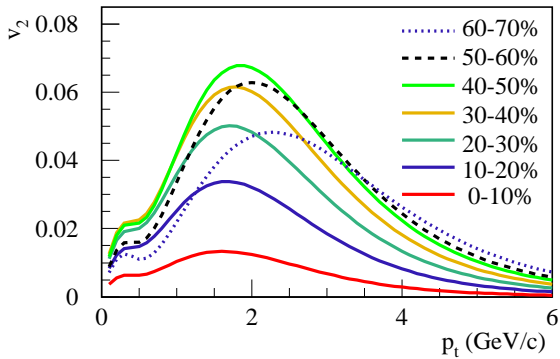}

\caption{\label{v2} (Color online) Midrapidity $v_{2}$ for thermal photons in Au+Au
collisions at $\sqrt{s_{NN}}=200$ GeV is shown for various centralities
from 0 to 70\% in $0<\pt<6$ GeV/$c$. }
\end{figure}

In Fig.~\ref{v2}, the $v_{2}$ of thermal photons at midrapidity $y=0$ in $0<\pt<6$
GeV/$c$ is shown for various centralities from 0 to 70\% in Au+Au
collisions at $\sqrt{s_{NN}}=200$ GeV. The solid lines from bottom
to top refer to centralities 0-10\%, 10-20\%, 20-30\%, 30-40\%, and
40-50\%, respectively. The dashed lines from top to bottom at $\pt=$
2 GeV/$c$ refer to the centralities 50-60\% and 60-70\%.

For each centrality, the thermal $v_{2}$ increases then decreases
with increasing $p_{t}$ and a peak appears at $p_{t}\sim2$~GeV/$c$.
This $p_{t}$ dependence is consistent with the prediction based on 
2+1D hydrodynamics\cite{ther} and explained as the weak transverse 
flow at the early stage. 

\begin{figure}
\includegraphics[scale=0.8]{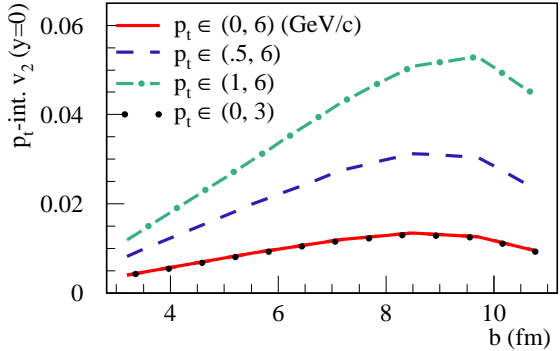}

\caption{\label{v2int} (Color online)  The  $\pt$-integrated $v_{2}$ at $y=0$ is plotted as a function of impact parameter $b$. The dependence of $\pt$-integrated range is presented with  various  types of curves.     }
\end{figure}

The centrality dependence is shown clearly in Fig.~\ref{v2int}, where the $\pt$-integrated $v_2$ at $y=0$ is plotted as a function of impact parameter $b$. 
The $\pt$-integrated $v_{2}$ reaches maximum at about 50\% centrality,
due to the interplay between the asymmetry and the strength of the transverse flow,
as shown in Tab.I. 

In Fig.~\ref{v2int}, the dependence of $\pt$-integrated range is presented with  various types of curves.  The  $\pt$-integrated $v_{2}$  is not sensitive to the upper limit, as one can see that the curves from integrated range (0,3) and (0,6)GeV do not differ. But it is very sensitive to the lower limit, as shown in  Fig.~\ref{v2int}, where three lower limits of $\pt$-integral, 0, 0.5 and 1GeV/c, are used.  
The reason is that the $\pt$ spectrum of thermal photons decreases so rapidly\cite{Liu:2008eh} that the  $v_2$ weight at low $\pt$ spectrum plays an very important role.

\section{Conclusion} 
Based on a three-dimensional ideal hydrodynamics,
we calculated the elliptic flow of thermal photons $v_{2}$ 
from Au+Au collisions at $\sqrt{s_{NN}}=200$ GeV.  
Due to the interplay between the asymmetry and the strength of the transverse flow, thermal photon $v_{2}$ reaches maximum at  $\pt \sim $  2GeV/$c$ and  $\pt$-integrated $v_{2}$ reaches maximum at about 50\% centrality.

The insensitivity of the $\pt$-integrated $v_{2}$ to upper $\pt$-integral limit is very useful. In fact if we include all sources of direct photons, the $\pt$-integrated $v_{2}$ will not differ much, for two reasons:\\
1) The yield of direct photons at higher $\pt$ is  small because the $\pt$ spectrum of direct photons decreases rapidly though all sources of direct photons are counted\cite{Liu:2008eh}.  \\
2) The $v_2$  is also very small at higher $\pt$ for obvious reason that the main contribution at high $\pt$, the leading order contribution from nucleon-nucleon collisions, has a vanishing $v_2$. \\
So we can predict the  $\pt$-integrated  $v_{2}$ are dominated by thermal photons.

The $\pt$-integrated $v_{2}$ is very sensitive to the lower integral limit.
Experimentally, the smallest measurable $\pt$ is up to the detector.
Above this value, experimentalists can still vary the lower integral limit of  $\pt$
to check how reliable the photon emission rate formula is. Theoretically this formula is very close to non-perturbative region and not reliable.

Though 3D hydrodynamics provides similar results as from 2D hydrodynamics
at midrapidity, the rapidity dependence will be certainly different,
which will be presented later. 

\section*{Acknowledgments}
This work is supported by the Natural Science Foundation of China
under the project No. 10505010 and MOE of China under project No.~IRT0624.
The work of T.H. was partly supported by Grant-in-Aid for Scientific
Research No.~19740130 and by Sumitomo Foundation No.~080734. FML
is particularly grateful to U.~Heinz, R.~Chatterjee and D.~K.~Srivastava,
for the helpful discussion.

\end{document}